\documentclass[journal,10pt]{IEEEtran}

\usepackage{amssymb}
\usepackage{amsmath}
\usepackage{cite}
\usepackage{url}
\usepackage{xcolor}
\usepackage{cite,graphicx,amsmath,amssymb}
\usepackage{subfigure}
\usepackage{citesort}
\usepackage{fancyhdr}
\usepackage{mdwmath}
\usepackage{mdwtab}
\usepackage{caption}
\usepackage{amsthm}

\newtheorem{remark}{Remark}

\newtheorem{theorem}{Theorem}

\newtheorem{lemma}{Lemma}

\newtheorem{corollary}{Corollary}

\newtheorem{assumption}{Assumption}

\newtheorem{definition}{Definition}

\makeatletter
\def\ScaleIfNeeded{%
\ifdim\Gin@nat@width>\linewidth \linewidth \else \Gin@nat@width
\fi } \makeatother

\begin{document}

\title{Signal Fractions Analysis and Safety-Distance Modeling in V2V Inter-lane Communications}

\author{Wenqiang~Yi, Yuanwei~Liu, and Arumugam~Nallanathan
\thanks{W. Yi, Y. Liu, A. Nallanathan are with Queen Mary University of London, London E1 4NS, U.K. (email: \{w.yi, yuanwei.liu, a.nallanathan\}@qmul.ac.uk).}
}

\maketitle

\begin{abstract}
  For vehicular networks, safety distances are important, but existing spatial models fail to characterize this parameter, especially for inter-lane communications. This work proposes a Mat\'ern hard-core processes based framework to appraise the performance of signal fractions (SF), where the hard-core distance is used to depict safety distances. By considering both semicircle and omnidirectional antennas, we derive high-accurate closed-form probability density functions of communication distances to acquire the complementary cumulative distribution function of SF. The derived expressions theoretically demonstrate that the nearest vehicle within the safety distance follows a uniform distribution and there is an upper limit for SF in terms of the transmit power.
\end{abstract}


\section{Introduction}
With the rapid development of self-driving vehicles, in addition to sensing techniques, vehicle-to-vehicle (V2V) communications for transmitting instant instructions among vehicles become indispensable~\cite{7999286}. To evaluate the performance of large-scale V2V networks, practical spatial models are essential but challenging. Stochastic geometry is an efficient mathematical tool to model the locations of devices in wireless networks~\cite{illian2008statistical}. The authors in~\cite{7331645} illustrated that multiple 1D PPPs have the similar performance to 2D-PPP approaches. To characterize the safety distance between vehicles, the authors in~\cite{8115190} attempted Mat\'ern hard-core processes (MHCP) to model the vehicles. For simplicity, the MHCP is regarded as a stationary thinning PPP. By utilizing a dynamic thinning PPP, the evaluation accuracy was improved in a recent work~\cite{8876629}. However, \cite{8876629} omits a vital scenario, namely inter-lane communications, which is important for the lane-changing negotiation, intra-lane see-through ability, emergency/congestion broadcasting, etc. Such omission motivates this work.

Regarding evaluation metrics, existing performance-analysis works focus on coverage probabilities (outage probabilities)~\cite{7999286,8876629,8115190}. One main shortage of coverage probabilities is that the value range of the independent variable, namely the signal-to-interference-plus-noise ratio (SINR), is infinite. It is difficult to evaluate the overall trend of this metric, especially for numerical results. For V2V communications requiring ultra-high quality of services, the overall trend is a key factor for appraising technical designs since hidden drawbacks may be contained in the omitted value range. Therefore, this paper uses a new metric with a finite value range, namely signal fractions (SF)~\cite{haenggi2020sir}, which helps to show the complete information in a single numerical figure. Due to the finite value range, the integrals based on SF, e.g., the moments, do not turn to infinity. Note that the complementary cumulative distribution function (CCDF) of SF equals coverage probabilities but with different independent variables.

The main contributions of this work are: 1) We design a spatial model for intra-lane V2V communications based on multiple 1D MHCPs, where both the semicircle and omnidirectional antennas are considered; 2) Closed-form probability density functions (PDF) of the inter-lane communication distance are obtained; 3) A tractable CCDF of SF is provided based on a dynamic thinning PPP that has higher accuracy than the traditional method with a stationary thinning PPP.

\section{System Model}\label{System Model}
\begin{figure}[t!]
  \centering
  \includegraphics[width= 3.5 in]{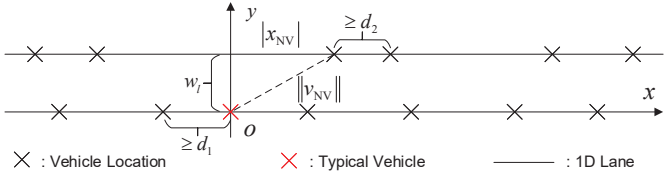}
  \caption{Illustration of the spatial model for the considered V2V communications.}\label{fig0}
\end{figure}

As shown in Fig.~\ref{fig0}, a road with two 1D lanes is considered in this work, where vehicles in the $i$-th lane $(i\in\{1,2\})$ are modeled according to a 1D MHCP $\Phi_i\subset \mathbb{R}$ with a hard-core distance $d_i$~\cite{illian2008statistical}. The generating PPP for all MHCPs is $\Phi_p$ with a density $\lambda_p$. A randomly selected typical vehicle in the first lane is fixed at the origin. Considering the distance between two lanes $w_l$, the 2D locations of all vehicles $\mathbb{U}_v \subset \mathbb{R}^2$ is
\begin{align}
\mathbb{U}_v = \bigcup\limits_{i=1}^{2} \bigcup\limits_{x_i\in\Phi_i} \{\mathbf{u} (x_i, (i-1)w_l)\},
\end{align}
where $\mathbf{u} (x,y)$ represents one point in the Cartesian coordinate system. Comparing with~\cite{8876629} that considers the range $[0,\infty)$ for $x$ coordinates, this work assumes that $x \in (-\infty,\infty)$ to provide a general spatial model, which is vital for omnidirectional antennas. The definition of MHCP is provided as follows.

\begin{definition}\label{def1}
(Mat{\'e}rn hard-core process): A type II MHCP $\Phi_h$ is generated from a generating PPP $\Phi_g$. Given a mark $m_\mathbf{x}$, which is uniformly distributed in $[0,1]$, to each point $\mathbf{x} \in \Phi_g$, $\Phi_h$ retains the point $\mathbf{x}$ which obeys $m_\mathbf{x} > m_{\mathbf{x}'}, \forall \mathbf{x}' \in \mathbb{S}(\mathbf{x}) \cap \Phi_g $. The $\mathbb{S}(\mathbf{x})$ is the 1D ball centered at $\mathbf{x}$ with radius equalling to the hard-core distance.
\end{definition}

From the definition, it worth noting that the distance between any two points in $\Phi_i$ is larger than $d_i$. Assuming the length of a vehicle is $d_v$ and the safety distance between two vehicles is $d_s$\footnote{The safety distance is decided by the speed of vehicles denoted by $v_s$. Following a two-second rule in driving, we have $d_s=2v_s$.}, we define that $d_i \equiv d_v + d_s$. According to~\cite{8876629}, the first order density of $\Phi_i$ is $\lambda_i= \frac{1-\exp(-2\lambda_pd_i)}{2d_i}$ and the corresponding second order density is
\begin{align}
{\lambda_i^{(2)}}(r) &= \left\{ {\begin{array}{*{20}{l}}
   {0,} & {0< r \le {d_i}}  \\
   {\frac{2\lambda_i}{ r }-\frac{2 (1- e^{-\lambda_p(2d_{i}+r)})}{ r (2d_{i}+r)},} & {{d_i} \le r < 2{d_i}}.  \\
   {\lambda_i^2,} & {r \ge 2{d_i}}  \\
\end{array}} \right.
\end{align}
\subsection{Association Scheme}
The V2V communication in the same lane has been investigated in~\cite{8876629}. This work considers another user association scheme: the nearest vehicle (NV) scheme, where the typical vehicle connects to the nearest vehicle in the adjacent lane.

For the NV scheme, the y-coordinate of the serving vehicle is $y_{\mathrm{NV}}=w_l$ and the x-coordinate can be expressed as
\begin{align}
x_{\mathrm{NV}} = \arg \min\limits_{x_2\in \Phi_2} |x_2|.
\end{align}
Therefore the location of the serving vehicle is $\mathbf{u}_{\mathrm{NV}} (x_{\mathrm{NV}},y_{\mathrm{NV}})\in \mathbb{U}_v$.

The condition $({0< x_{\mathrm{NV}} \le {d_2}})$ represents \emph{a practical scenario of lane changing}. For example, when the in front serving vehicle wants to join to the typical vehicle's lane, it checks the horizontal distance $x_{\mathrm{NV}}$ first. If the space between them is less than the safety distance, namely ${0< x_{\mathrm{NV}} \le {d_2}}$, it needs to transmit its speed and acceleration to the typical vehicle to let the typical vehicle to do corresponding actions.
\subsection{Signal Model}
We consider the metric SF instead of SINR. Since SINR is in a infinite range $\mathbb{R}^+$, while SF is in a finite range, i.e., $[0,1)$, it is simpler to exploit the entire distribution of the SF. The definition of SF is~\cite{haenggi2020sir}
\begin{definition}\label{def2}
(Signal fraction): The SF is the ratio of the signal power $S$ to the total received power including interference $I$ and noise $N$, which can be expressed as
\begin{align}
\mathrm{SF}  \buildrel \Delta \over =  \frac{S}{S+I+N} = \frac{\mathrm{SINR}}{\mathrm{SINR}+1}.
\end{align}
\end{definition}
Based on \textbf{Definition~\ref{def2}}, the CCDF of SINR $\bar{F}_{\mathrm{SINR}}$ can be calculated with the aid of the CCDF of SF $\bar{F}_{\mathrm{SF}}$, i.e.,
\begin{align}\label{change}
\bar{F}_{\mathrm{SF}}(\sigma) = \bar{F}_{\mathrm{SINR}}\Big(\frac{\sigma}{1-\sigma}\Big).
\end{align}

To enhance the generality, we consider two cases of antennas with unit antenna gain. \textbf{Case 1}: A semicircle antenna is deployed at the top of each vehicle. When transmitting, the antenna faces the behind, while the direction is changed to the front when receiving. \textbf{Case 2}: An omnidirectional antenna is used instead. After that, the SF can be expressed as
\begin{align}
\mathrm{SF}_\chi = \frac{|h_{\mathbf{u}_{\mathrm{NV}}}|^2 \|\mathbf{u}_{\mathrm{NV}}\|^{-\alpha}}{\sum\limits_{\mathbf{u}\in \mathbb{U}^\chi_v} |h_{\mathbf{u}}|^2 \|\mathbf{u}\|^{-\alpha}+\rho},
\end{align}
where $\chi\in\{c_1,c_2\}$ is the case indicator and $c_1/c_2$ represents the \textbf{Case 1}/\textbf{Case 2}. The set $\mathbb{U}^{c_1}_v = \mathbb{U}_v|_{x_i>0}$ and $\mathbb{U}^{c_2}_v = \mathbb{U}_v$. The $\rho = \frac{N}{P_tC}$, where $P_t$ is the transmit power for all vehicles, $C$ is the free space path loss with a reference distance $d_0$, and $\alpha$ is the path loss exponent. The $h_{\mathbf{u}}$ represents the Rayleigh fading term for the link with the transmitter at ${\mathbf{u}}$.
\section{Performance Evaluation}
For light traffic, namely $\lambda_p d_i \to 0$, $\Phi_i \approx \Phi_p$, which is well-investigated in existing researches. Therefore, this work only discuss heavy traffic scenarios $(\lambda_p d_i \geq 1)$. Before analyzing the SF performance, we first derive the distribution of the distance between the typical vehicle and its serving vehicle.
\subsection{Distance Distributions}
Under the NV scheme, the typical vehicle and the serving vehicle are respectively located in the first and second lane.
\begin{lemma}\label{lemma1}
\emph{For the serving vehicle located in the adjacent lane, the PDF of the horizontal distance $x_{\mathrm{NV}}$ under \textbf{Case 1} is}
\begin{align}
{f^{c_1}_x}\left( {{r_1}} \right) \approx \left\{ {\begin{array}{*{20}{l}}
   {{\lambda_2} ,} & {{0< r_1} \le {d_2}}  \\
   {{\lambda_2}( {1 - {\lambda_2}( {{r_1} - {d_2}} )} ),} & {{d_2} < {r_1} \le 2{d_2}},  \\
   {{\lambda _r}\exp \left( { - {\lambda _r}r_1} \right),} & {{r_1} > 2{d_2}}  \\
\end{array}} \right.
\end{align}
\emph{where $\lambda_r = \ln\left(\frac{2}{{{{\left( {{\lambda_2}{d_2} - 2} \right)}^2-2}}}\right)/(2d_2)$.}
\begin{IEEEproof}
See Appendix A.
\end{IEEEproof}
\end{lemma}
\begin{remark}
From \emph{\textbf{Lemma~\ref{lemma1}}}, it can be concluded that the distribution of $x_{\mathrm{NV}}$ in the range $(0,d_2]$ is a uniform distribution. Therefore, the PDF of the practical scenario of lane changing is ${f_x^{c_1}}\left( {{r_1}}| {{0< r_1} \le {d_2}} \right) = 1/d_2$.
\end{remark}

For omnidirectional antennas, the serving and interfering vehicles can be either from the front or the back, namely $x_{\mathrm{NV}}$ can be smaller than zero. Under this case, we have the following corollary.
\begin{corollary}
\emph{Base on \textbf{Lemma~\ref{lemma1}}, the PDF of the horizontal distance $|x_{\mathrm{NV}}|$ under \textbf{Case 2} is given by}
\begin{align}
{f^{c_2}_x}\left( {{r_1}} \right) \approx \left\{ {\begin{array}{*{20}{l}}
   {2{\lambda _2},} & {0 < {r_1} \le \frac{{{d_2}}}{2}}  \\
   {g\left( {{r_1}} \right),} & {\frac{{{d_2}}}{2} < {r_1} \le \frac{{3{d_2}}}{2}}  \\
   {\frac{{2{\lambda _r}\exp \left( { - {\lambda _r}\left( {2{r_1} + d} \right)} \right)}}{{1 - {\lambda _2}{d_2}}},} & {{r_1} > \frac{{3{d_2}}}{2}}  \\
\end{array}} \right.,
\end{align}
\emph{where $g(r_1) = \frac{{2\lambda_2( {1 + {\lambda _2}{d_2}/2 - {\lambda _2}{r_1}} )}}{{1 - {\lambda _2}{d_2}}}\big( {\frac{{3{\lambda _2}{d_2}}}{2} - \frac{{3\lambda _2^2d_2^2}}{8} + e^{- 2{\lambda _r}{d_2} }}$ $- \big( {{\lambda _2} + \frac{{\lambda _2^2{d_2}}}{2}} \big){r_1} + \frac{{\lambda _2^2r_1^2}}{2} \big)$.}
\begin{IEEEproof}
In this case, $x_{\mathrm{NV}}\in(-\infty,\infty)$. Based on ${f^{c_1}_x}\left( {{r_1}} \right)$, $x_{\mathrm{NV}}\in[-\frac{d_2}{2},\frac{d_2}{2}]$ is uniformly distributed with a density $\lambda_2$. For $x_{\mathrm{NV}}\in(-\infty,-\frac{d_2}{2}] \cup [\frac{d_2}{2},\infty)$, we ignore the negligible correlation between the point process in $(-\infty,-\frac{d_2}{2}]$ and $[\frac{d_2}{2},\infty)$. Note that the CDF of $Y=\min(X_1,X_2)$ is $F_Y = \Pr[\min(X_1,X_2)<x]=1-(1-F_X(x))^2$, where $F_X$ is the CDF of $X_1$ and $X_2$. After several algebraic manipulations, we obtain this corollary.
\end{IEEEproof}
\end{corollary}

Regarding the communication distance $\|\mathbf{v}_\mathrm{NV}\|$ between the serving vehicle and the typical vehicle, since the serving vehicle is in the second lane, the vertical distance is $w_l$. According the Pythagorean theorem, this distance is $\|\mathbf{v}_\mathrm{NV}\|=\sqrt{|x_{\mathrm{NV}}|^2+w_l^2}$.
\begin{corollary}
\emph{Based on the NV scheme, the PDF of the communication distance $\|\mathbf{v}_\mathrm{NV}\|> w_l$ can be expressed as follows}
\begin{align}
f^{\chi}_r(r)=\frac{r}{\sqrt{r^2-w_l^2}}f_x^\chi\Big(\sqrt{r^2-w_l^2}\Big),
\end{align}
where $\chi\in\{c_1,c_2\}$ is the case indicator.
\begin{IEEEproof}
The CDF of $\|\mathbf{v}_\mathrm{NV}\|$ is $F^{\chi}_r(r)=\Pr[\sqrt{|x_{\mathrm{NV}}|^2+w_l^2}<r]=\Pr[|x_{\mathrm{NV}}|^2<\sqrt{r^2-w_l^2}]=F^{\chi}_x(\sqrt{r^2-w_l^2})$. Based on $f^{\chi}_r(r)=\frac{dF^{\chi}_r(r)}{dr}=\frac{dF^{\chi}_x(\sqrt{r^2-w_l^2})}{dr}$, we have this corollary.
\end{IEEEproof}
\end{corollary}
\subsection{Signal Fraction Performance}
Given a target data rate $R_{t}$, the target SINR can be derived via Shannon-Hartley theorem, that is $\gamma_t = 2^{R_t/B}-1$, where $B$ is the system bandwidth. Before evaluating the performance of SF, we first derive the CCDF of SINR, namely the coverage probability, which is defined as
\begin{align}\label{cp}
\bar{F}^{\chi}_{\mathrm{SINR}}(\gamma_t) = \Pr\Big[\frac{|h_{\mathbf{u}_{\mathrm{NV}}}|^2 \|\mathbf{u}_{\mathrm{NV}}\|^{-\alpha}}{\sum\limits_{\mathbf{u}\in {\mathbb{U}^\chi_v}\setminus {\mathbf{u}_{\mathrm{NV}}} } |h_{\mathbf{u}}|^2 \|\mathbf{u}\|^{-\alpha}+\rho}>\gamma_t\Big]
\end{align}
\begin{theorem}\label{theorem1}
\emph{Under the NV scheme, the coverage probability for inter-lane communications between two adjacent lanes is}
\begin{align}\label{result}
\bar{F}^{\chi}_{\mathrm{SINR}}(\gamma_t) \approx \int_{w_l}^\infty  {\exp ( { - {\gamma _t}( {\rho  + \beta_{\chi}({I_1} + {I_2}}) )r^\alpha } )} {f^{\chi}_r}( {{r}} )d{r},
\end{align}
\emph{where $I_1 = \lambda _1^{ - 1}\int_{{d_1}}^{ \infty } {\lambda _1^{\left( 2 \right)}} {r_d^{ - \alpha }}dr_d$, ${I_2} = \int_0^\infty  {\int_{{d_2}}^{ \infty } {\frac{{\lambda _2^{( 2 )}\left( r \right){f^{c_1}_x}(r_1)}}{{{{\lambda _2( {{{( {r_1 + {{r_d}} } )}^2} + w_l^2} )}^{\alpha /2}}}}dr_d d{r_1}} }$, $\beta_{c_1} = 1$, and $\beta_{c_2} = 2$.}
\begin{IEEEproof}
Since $|h_{\mathbf{u}_{\mathrm{NV}}}|^2$ follows the $\exp(1)$ distribution, \eqref{cp} can be rewritten as
\begin{align}\label{process}
&\bar{F}^\chi_{\mathrm{SINR}}(\gamma_t) =\mathbb{E}\Big[\exp\Big(-\gamma_t\Big(\sum\limits_{\mathbf{u}\in {\mathbb{U}^\chi_v}\setminus {\mathbf{u}_{\mathrm{NV}}} } \frac{|h_{\mathbf{u}}|^2} {\|\mathbf{u}\|^{\alpha}}+\rho\Big)r^\alpha\Big)\Big]\nonumber \\
&\mathop  \approx \limits^{(a)} \int_{w_l}^\infty e^{\mathbb{E}\Big[-\gamma_t\Big(\sum\limits_{\mathbf{u}\in {\mathbb{U}^\chi_v}\setminus {\mathbf{u}_{\mathrm{NV}}} } \frac{|h_{\mathbf{u}}|^2} {\|\mathbf{u}\|^{\alpha}}+\rho\Big)r^\alpha\Big]}f^\chi_r(r)dr,
\end{align}
where $(a)$ uses Jensen's inequality. For the interference under \emph{\textbf{Case 2}}, we ignore the negligible correlation between the point distributed in $(-\infty,0)$ and $(0,\infty)$, so $\beta_{c_2} = 2\beta_{c_1} = 2$. By applying Campbell's theorem~\cite{daley2007introduction} into~\eqref{process}, we obtain~\eqref{result}.
\end{IEEEproof}
\end{theorem}
\begin{remark}\label{remark2}
From \emph{\textbf{Theorem~\ref{theorem1}}}, we find the coverage probability is inversely proportional to $\rho$. The upper limit for $\bar{F}^\chi_{\mathrm{SINR}}$ is $\bar{F}^\chi_{\mathrm{SINR}}|_{\rho = 0}$.
\end{remark}

It is worth noting that when $\gamma _t$ is small, $\bar{F}^\chi_{\mathrm{SINR}}(\gamma_t)$ can be approximated by $F^\chi_1(\gamma _t) = \int_{w_l}^\infty  { ( 1{ - {\gamma _t}( {\rho  + \beta_\chi({I_1} + {I_2})} )r^\alpha } )}{f^\chi_r}( {{r}} )d{r}$ since $\lim\limits_{x\to 0}\exp(-x) = 1-x $. When $\gamma _t$ is large, it can be approximated by $F^\chi_2(\gamma _t)=\lambda_2\int_{w_l}^{\sqrt{d_2^2+w_l^2}} \frac {r\exp ( { - {\gamma _t}( {\rho  + \beta_\chi({I_1} + {I_2})} )r^\alpha } )}{\sqrt{r^2-w_l^2}} d{r}$ since the uniform-distribution part dominates $\bar{F}^\chi_{\mathrm{SINR}}(\gamma_t)$.

\begin{corollary}\label{corollary1}
\emph{Based on the CCDF of SINR, the corresponding CCDF of SF is given by}
\begin{align}
\bar{F}^\chi_{\mathrm{SF}}(\sigma) \approx \int_{w_l}^\infty  {\exp \Big( { - \frac{\sigma( {\rho  + {I_1} + {I_2}} )r_1^\alpha}{1-\sigma} } \Big)} {f^{\chi}_x}(r_1)d{r_1}.
\end{align}
\begin{IEEEproof}
By applying \eqref{change} to \emph{\textbf{Theorem~\ref{theorem1}}}, we obtain this corollary.
\end{IEEEproof}
\end{corollary}
The $\bar{F}^\chi_{\mathrm{SF}}(\sigma)$ has the same property as discussed in \textbf{Remark~\ref{remark2}}. Moreover, $\bar{F}^\chi_{\mathrm{SF}}(\sigma)$ can also be approximated using the same method as for $\bar{F}^\chi_{\mathrm{SINR}}(\gamma_t)$.
\section{Numerical Result}
As introduced in~\cite{haenggi2020sir}, we apply the M\"obius homeomerphic (MH) unit to evaluate performance of SF, i.e., $\sigma$ MH = $\frac{\sigma}{1-\sigma}$, $(\sigma\in[0,1))$. The network setting is listed as follows: $C=\frac{\lambda_w^2}{16\pi^2 d_0^2}$, the wavelength for $5$ GHz is $\lambda_w = 3\times 10^8 / (5\times 10^9)=6$ cm, the reference distance $d_0=1$ m, $\alpha =4$, $N=-90$ dBm, $P_t = 30$ dBm, $\lambda_p=1/10$ m$^{-1}$, $d_v=5$ m, and $w_l=5$ m.
\begin{figure}[t!]
  \centering
  \includegraphics[width= 3.5 in]{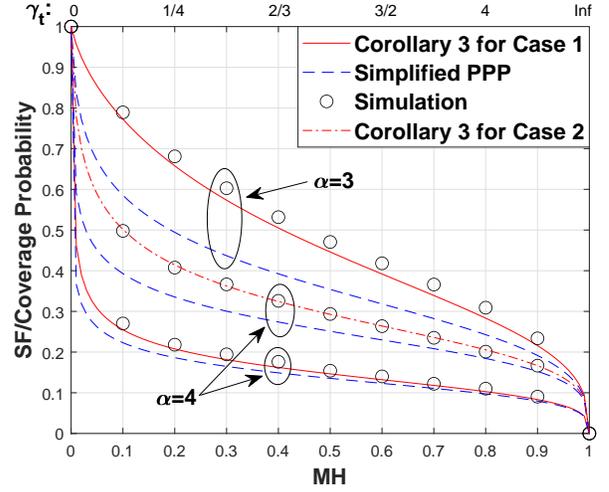}
  \caption{SF (Coverage probability) versus $\sigma$ in MH ($\gamma_t$), with $d_s=145$ m and the comparison with a same density PPP~\cite{8115190}.}\label{fig1}
\end{figure}
\begin{figure}[t!]
  \centering
  \includegraphics[width= 3.5 in]{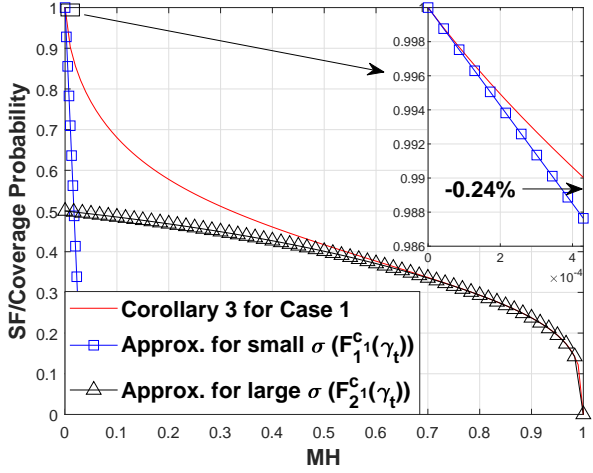}
  \caption{Validating of approximation expressions, with $\lambda_p=1/5$ m$^{-1}$ and $d_s=45$ m.}\label{fig2}
\end{figure}
\begin{figure}[t!]
  \centering
  \includegraphics[width= 3.5 in]{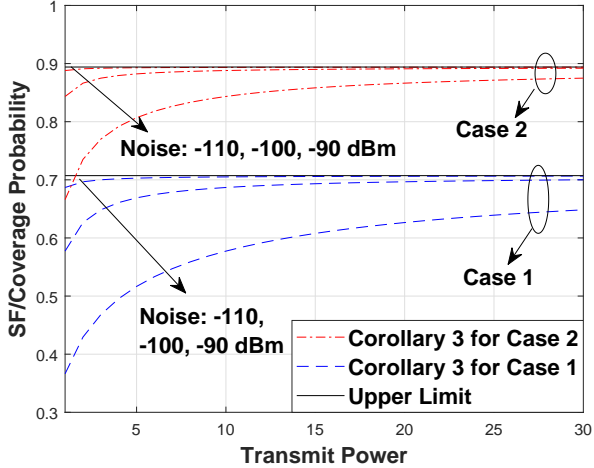}
  \caption{SF (Coverage probability) versus transmit power in W, with $d_s=95$ and $\sigma=1/2$ MH.}\label{fig3}
\end{figure}

Fig.~\ref{fig1} shows the difference between the proposed method and the traditional method, which uses a replacement PPP with density $\lambda_i$ to evaluate a MHCP~\cite{8115190}. Compared with the Monte Carlo simulations, the proposed CCDF of SF has a higher accuracy than the replacement PPP method, especially when $\alpha$ is small. Moreover, the omnidirectional antenna outperforms the semicircle antenna under the considered scenario. Note that when $\alpha$ is large, the near vehicles with short communication distance dominant the V2V network. Since the accuracy of the derived PDFs decreases with the increase of communication distances, our methods have higher accuracy in large $\alpha$ regimes than small $\alpha$ regimes. Fig.~\ref{fig1} also illustrates that SF explores the entire value range, i.e., $[0,1)$ with the finite independent variable $\sigma \in [0,1)$, while the independent variable for $\bar{F}^\chi_{\mathrm{SINR}}(\gamma_t)$ has an infinite value range, namely $\gamma_t \in [0,\inf)$.

\begin{figure*}[t]
\normalsize
  \begin{align}\label{A2}
 & {f^{c_1}_x}(r_1)= \Pr\big[m_{\mathbf{v}}>m_{\mathbf{v}'}, \mathbf{v} \in \Phi_p\cap \mathbb{L}(0,d_2), \mathbf{v}' \in \Phi_p\cap \mathbb{L}\left(r_1-d_2,r_1+d_2\right) \setminus \mathbf{v}\big]\nonumber\\
  \mathop  = \limits^{(a)} &\Pr\big[m_{\mathbf{v}}>m_{\mathbf{v}'_1}, \mathbf{v} \in \Phi_p\cap \mathbb{L}(0,d_2), \mathbf{v}'_1 \in \Phi_p\cap \mathbb{L}(0,d_2) \setminus \mathbf{v}\big]\Pr\big[m_{\mathbf{v}}>m_{\mathbf{v}'_2}, \mathbf{v}'_2 \in \Phi_p\cap \left(\mathbb{L}(r_1-d_2,0) \cup \mathbb{L}(d_2,r_1+d_2)\right)\big]\nonumber \\
   \mathop  = \limits^{(b)} & \sum\limits_{n = 1}^\infty  {\frac{{{{\left( {{\lambda _p}{d_2}} \right)}^n}}}{{n!}}\exp \left( { - {\lambda _p}{d_2}} \right)} {n \choose 1}\frac{1}{{{d_2}}}\int_0^1 {m_\mathbf{v}^{n - 1}\exp \left( { - \left( {1 - {m_\mathbf{v}}} \right){\lambda _p}{d_2}} \right)d{m_\mathbf{v}}}   \mathop  = \limits^{(c)}  \frac{{1 - \exp \left( { - 2{\lambda _p}{d_2}} \right)}}{{2{d_2}}} = {\lambda_2}.\tag{A.2}
\end{align}
\hrulefill \vspace*{0pt}
\end{figure*}

Fig.~\ref{fig2} validates the approximation expressions for the CCDF of SF. We study \textbf{Case 1} here as an example. When the value of $\sigma$ is small, $\bar{F}^{c_1}_{\mathrm{SF}}(\sigma)\approx F_1\left(\frac{\sigma}{1-\sigma}\right)$. Under the considered condition, the approximation error for $\bar{F}^{c_1}_{\mathrm{SF}}(\sigma) = \bar{F}^{c_1}_{\mathrm{SINR}} (\gamma_t) = 99\%$ is around $-0.24\%$. If unmanned vehicles require $\bar{F}^{c_1}_{\mathrm{SINR}} (\gamma_t)\ge 99\%$, namely the outage probability is less than $1\%$, $F^{c_1}_1\left(\gamma_t\right)$ can be utilized for simplicity. On the other hand, when the value of $\sigma$ is large, $\bar{F}^{c_1}_{\mathrm{SF}}(\sigma)\approx F^{c_1}_2\left(\frac{\sigma}{1-\sigma}\right)$.

Fig.~\ref{fig3} demonstrates the proposed insight. As discussed in \textbf{Remark~\ref{remark2}}, Fig.~\ref{fig3} shows that $\bar{F}^{\chi}_{\mathrm{SF}}$ is proportional to the transmit power $P_t$ and inversely proportional to the noise power for both cases. Moreover, Fig.~\ref{fig3} also illustrates that there is an upper limit of $\bar{F}^{\chi}_{\mathrm{SF}}$, which means if the required coverage probability exceeds this upper limit, changing transmit power cannot satisfy the demand and hence additional interference cancellation methods are needed.
\section{Conclusion}
This work has studied the SF performance for inter-lane V2V communications with the aid of MHCPs. Closed-form PDFs for communication distances have been provided. Based on these PDFs, the tractable CCDF of SF has been derived as well as several tight approximation expressions. Since there exists an upper limit of the SF performance, our future work will study interference cancellation techniques for dense V2V communications to break this limit.

\numberwithin{equation}{section}
\section*{Appendix~A: Proof of Lemma~\ref{lemma1}}
\renewcommand{\theequation}{A.\arabic{equation}}
\setcounter{equation}{0}
Under \textbf{Case 1}, we discuss the distribution of $x_{\mathrm{NV}}$ in three ranges: $(0,d_2]$, $(d_2,2d_2]$, and $(2d_2,\infty)$.
\subsubsection{$0<x_{\mathrm{NV}}\le d_2$}
The PDF of $x_{\mathrm{NV}}$ can be defined as
\begin{align}~\label{A1}
{f^{c_1}_x}(r_1) &= \Pr[\Phi_2\cap \mathbb{L}(0,d_2)= \mathbf{v}, \mathbf{v}\in \Phi_p], \tag{A.1}
\end{align}
where $\mathbb{L}(x_1,x_2)$ is the line segment from $x_1$ to $x_2$ in the second lane including the point at $x_2$. Based on the definition of MHCP, \eqref{A1} can be rewritten at the top of this page.

In \eqref{A2}, the process $(a)$ follows the fact that $\mathbb{L}\left(r_1-d_2,r_1+d_2\right)  = \mathbb{L}\left(r_1-d_2,0\right) \cup \mathbb{L}\left(0,d_2\right) \cup \mathbb{L}\left(d_2,r_1+d_2\right) $. $(b)$ considers the void probability of PPPs and the similar process of (10) in~\cite{7511676}. $(c)$ uses the power series of exponential function, i.e., $\exp(x)=\sum_{n=0}^\infty \frac{x^n}{n!}$.

\subsubsection{$d_2<x_{\mathrm{NV}}\le 2d_2$}
The PDF of $x_{\mathrm{NV}}$ can be defined as
\begin{align}
{f^{c_1}_x}(r_1) = &\Pr[\Phi_2\cap \mathbb{L}(d_2,2d_2)= \mathbf{v}, \mathbf{v}\in \Phi_p ] \nonumber \\
= &\Pr\big[\underbrace{\Phi_2\cap \mathbb{L}\left(r_1-d_2,r_1+d_2\right) = \mathbf{v}}_{\mathrm{C_1}}, \nonumber \\
&\ \ \ \ \ \underbrace{\Phi_2\cap \mathbb{L}(0,r_1-d_2) = \emptyset}_{\mathrm{C_2}}, \mathbf{v}\in \Phi_p \big] \nonumber \\
\mathop  \approx \limits^{(d)} &  \lambda_2(1-\lambda_2(r_1-d_2)),\tag{A.3}
\end{align}
where $(d)$ ignores the overlapping between the conditions $\mathrm{C_1}$ and $\mathrm{C_2}$. Assuming $\mathbf{v}_1\in \Phi_p \cap \mathbb{L}(0,r_1-d_2)$ and $\mathbf{v}_2\in \Phi_p \cap \mathbb{L}(r_1-d_2,r_1)$, this overlapping probability is
\begin{align}\label{A4}
P_o =& \Pr[m_{\mathbf{v}_1}<m_{\mathbf{v}_2}<m_{\mathbf{v}}, |{\mathbf{v}_2}-{\mathbf{v}_1}|<d_2]\nonumber \\
<&\Pr[m_{\mathbf{v}_1}<m_{\mathbf{v}_2}<m_{\mathbf{v}}]\nonumber \\
=&\int_0^1 \exp\left(-(1-m_{\mathbf{v}})\lambda_p d_2\right) d m_{\mathbf{v}} \nonumber \\
&\times \int_0^{m_{\mathbf{v}}} \exp\left(- (1-m_{\mathbf{v}_2})\lambda_p (r_1-d_2)\right) d m_{\mathbf{v}_2}. \tag{A.4}
\end{align}
Based on \eqref{A4}, when $\lambda_p d_2$ is large enough, we have $\lim\limits_{\lambda_p d_2 \to \infty} P_o \to 0$ and hence the overlapping is negligible.

\subsubsection{$x_{\mathrm{NV}}> 2d_2$}
Under this case, $x_{\mathrm{NV}}$ can be divided into multiple segments with length $d_2$ to calculate the PDF ${f^{c_1}_x}(r_1)$. For each segment, the derivation has the similar proof process for the case $(d_2<r_1\le 2d_2)$. However, this result delivers limited insights but with high complexity. Fortunately, when $\lambda_p d_2 \ge 1$, the CDF of $r_1$ in the range $[0,2d_2]$ is large enough:
\begin{align}
&{F_C}(2d_2)= \lambda_2d_2+\int_{d_2}^{2d_2} {\lambda_2\left( {1 - {\lambda_2}\left( {{r_1} - {d_2}} \right)} \right)dr_1}\nonumber \\
=& 2-\frac{(2-\lambda_2d_2)^2}{2} \ge 2-\frac{(3+\exp(-2))^2}{8} \approx 0.77.\tag{A.5}
\end{align}
Therefore, we use a replacement 1D PPP with a density $\lambda_r$ to approximate the PDF in the rest range, namely $[2d_2,+\infty)$. The density $\lambda_r$ obeys $\int_{2d_2}^{\infty} \lambda_r \exp(-\lambda_r r)dr = 1- {F_C}(2d_2)$. As a result, $\lambda_r = \ln\Big(\frac{2}{{{{\left( {{\lambda_2}{d_2} - 2} \right)}^2-2}}}\Big)/(2d_2)$.

The proof is completed.
\bibliographystyle{IEEEtran}
\bibliography{mybib}

\end{document}